\documentclass[10pt,a4paper,final]{iopart}
\usepackage{acronym}
\usepackage{bm}
\usepackage{enumerate}
\usepackage[top=2.2cm,bottom=2.3cm,left=2.9cm,right=2.9cm]{geometry}
\usepackage{graphicx}
\usepackage{hyperref}

\hypersetup{
pdftitle={Phase-locking an interferometer with single-photon detections},
pdfauthor={Bastian Hacker, Kevin Günthner, Conrad Rößler, Christoph Marquardt},
pdfkeywords = {interferometry, feedback, phase-locking, quantum communication, quantum key distribution, single-photon detection},
colorlinks=true, linkcolor=[RGB]{25,132,185}, citecolor=[RGB]{25,132,185}, filecolor=[RGB]{25,132,185}, urlcolor=[RGB]{25,132,185}}

\newacro{mzi}[MZI]{Mach-Zehnder interferometer}
\newacro{ou}[OU]{Ornstein-Uhlenbeck}
\newacro{pm}[PM]{Polarization-Maintaining}
\newacro{per}[PER]{Polarization Extinction Ratio}
\newacro{qkd}[QKD]{Quantum Key Distribution}
\newacro{SNSPD}[SNSPD]{superconducting nanowire single photon detector}

\DeclareTextFontCommand{\emph}{\textit}
\begin{document}
\title{Phase-locking an interferometer with single-photon detections}

\author{Bastian Hacker$^{1,2}$, Kevin Günthner$^{1,2}$, Conrad Rößler$^{1,2}$ and Christoph Marquardt$^{1,2}$}
\eads{\mailto{kevin.guenthner@mpl.mpg.de}, \mailto{christoph.marquardt@fau.de}}
\address{$^1$ Max Planck Institute for the Science of Light, Staudtstr. 2, 91058 Erlangen, Germany}
\address{$^2$ Friedrich-Alexander-Universität Erlangen-Nürnberg, Staudtstr. 7, A3, 91058 Erlangen, Germany}

\begin{abstract}
We report on a novel phase-locking technique for fiber-based Mach-Zehnder interferometers based on discrete single-photon detections, and demonstrate this in a setup. Our interferometer decodes relative-phase-encoded optical pulse pairs for quantum key distribution applications and requires no locking laser in addition to the weak received signal. Our new simple locking scheme is shown to produce an Ornstein-Uhlenbeck dynamic and achieve optimal phase noise for a given count rate. In case of wavelength drifts that arise during the reception of Doppler-shifted satellite signals, the arm-length difference gets continuously readjusted to keep the interferometer phase stable.
\end{abstract}


\submitto{\textit{New J. Phys.} \normalfont \textbf{25} 113007, doi: \href{https://doi.org/10.1088/1367-2630/ad0752}{10.1088/1367-2630/ad0752}}

\maketitle

\section{Introduction}
Quantum communication and specifically \ac{qkd} requires the encoding, transmission and reception of high-bandwidth signals with high fidelity. Encoding is possible in various degrees of freedom, typically polarization, time bin or phase \cite{xu_secure_2020}. The decoding of time bin and phase-encoded signals requires the interference of pulses from different time slots before measurement at chosen relative phases \cite{townsend_quantum_1994}. This is achieved with a phase-locked \ac{mzi}. Decoding of satellite \ac{qkd} signals poses additional challenges of a low signal level due to high propagation losses as well as a significantly varying Doppler shift \cite{lu_micius_2022}. Nevertheless, \ac{qkd} at loss levels above $50\,\mathrm{dB}$ is feasible \cite{bourgoin_experimental_2015} with the detection of single photons on modern \acp{SNSPD} that are available with high detection efficiency and timing precision down to few ps \cite{esmaeil_zadeh_superconducting_2021}.

Driven from these applications we pose the following question: How to optimize phase locking in the few-photon regime under realistic boundary conditions? In this work we investigate this question with an experimental setup and discuss the choice of optimal working points.

\begin{figure}[b]
    \centering
    \setlength{\unitlength}{15.3cm}
    \begin{picture}(1,0.34)
    \put(0,0){\includegraphics[width=8.2cm]{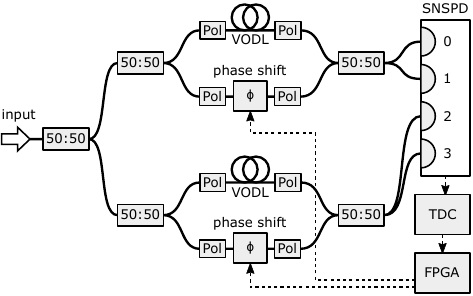}}
    \put(0.6,0){\includegraphics[width=6cm]{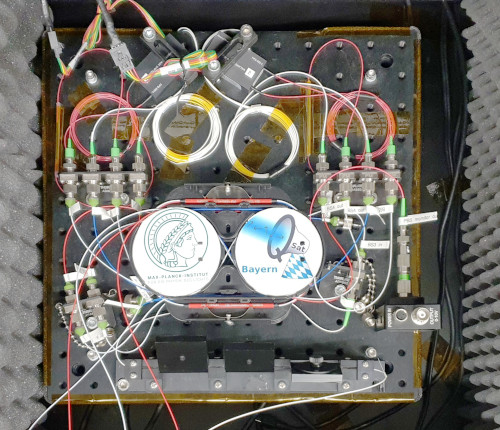}}
    \put(0,0.32){\normalsize(a)}
    \put(0.56,0.32){\normalsize(b)}
    \end{picture}
    \caption{(a) Schematic of the dual-fiber-\ac{mzi} setup with two independent locking phases. 50:50: beam splitter; Pol: In-fiber polarizer to counteract finite PER; VODL: variable optical delay line; phase shift: electrically controlled phase shifter; SNSPD: superconducting nanowire single-photon detector; TDC: time-to-digital converter; FPGA: field-programmable gate array. (b) Fiber-setup on breadboard in 19" rack drawer. Beam-splitters and stretchers are visible on the board.}
    \label{fig:interferometer+setup}
\end{figure}

\acp{mzi} consist of two subsequent beam splitters with two independent interferometer arms in between (\fref{fig:interferometer+setup}a). This configuration can decode phase information when two subsequent pulses of the incident signal get split into two different arms at the first beam splitter, then get delayed individually and finally interfere on the second beam splitter at a chosen relative phase. This directs light to an output port that depends on the incident relative pulse phase and thereby measures the phase.

The relative interferometer phase depends on the precise length of each two arms on a nanometer scale, and therefore requires active stabilization against random drifts \cite{jackson_elimination_1980, fritsch_simple_1981}. Conventionally optical interferometers are phase-locked with intensities in a range of nanowatts to watts, where the achievable accuracy is limited by the finite feedback-loop response time or mechanical actuator bandwidth, and not by photon shot-noise \cite{kirkendall_overview_2004}. In contrast, \ac{qkd} applications require signals on the level of single resolvable photons (femtowatts to picowatts), where an intense locking beam in the signal path is a huge disturbance. The issue is sometimes circumvented by using different wavelengths for the signal and locking \cite{cho_stabilization_2009, xavier_stable_2011, roztocki_arbitrary_2021, svarc_sub_2023}, which can be separated after the \ac{mzi}. Suppression of leakage from locking light into the signal path is however limited, and the signal phase becomes ambiguous after phase slips of the locking light.

To resolve this, the \acp{mzi} may be locked directly with the weak signal that is detected on single-photon detectors at count rates of kHz to MHz \cite{pulford_single_2005, yanikgonul_phase_2020}. Due to the gain-bandwidth product limit (connected to the fundamental Heisenberg number-phase uncertainty) \cite{zheng_ab_2019, muller_standard_2019}, low count rates allow only for slow feedback. At count rates of few kHz in \cite{pulford_single_2005, yanikgonul_phase_2020}, the resulting feedback bandwidth is in the Hz range. Thus, such a locking system cannot cancel acoustic noise in the kHz-range, where only few photons are received during one oscillation. Passive stability at those frequencies is therefore crucial \cite{micuda_highly_2014}. Low count rates call for an optimal use of the available information to reach the best achievable residual phase noise \cite{makarov_real-time_2004}. This work introduces such an optimal locking scheme, demonstrates the experimental implementation, and derives the achievable accuracy for any given system parameters.

\section{Setup}
\begin{figure}
\centering
    \setlength{\unitlength}{15.3cm}
    \begin{picture}(1,0.33)
    \put(0.02,0.06){\includegraphics[width=6cm]{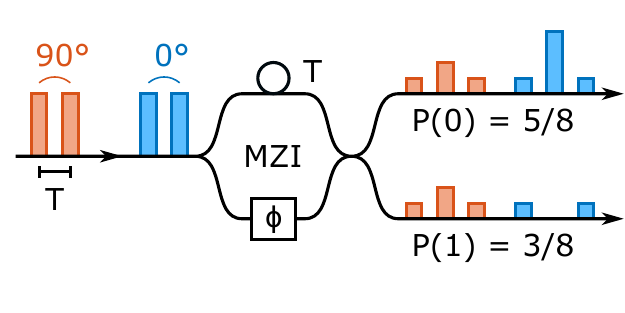}}
    \put(0.5,0){\includegraphics[width=7.5cm]{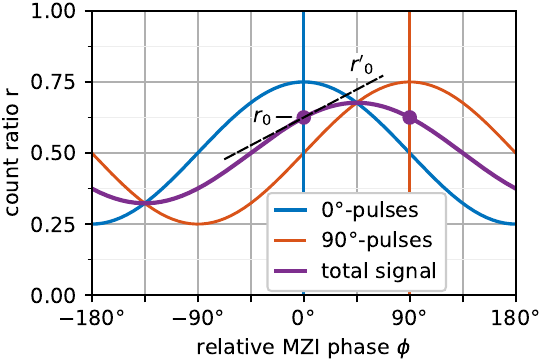}}
    \put(0,0.26){\normalsize(a)}
    \put(0.48,0.30){\normalsize(b)}
    \end{picture}
\caption{(a) Optical pulse pattern used for locking in our experiment. We receive alternating pulse pairs of $0^\circ$ and $90^\circ$ relative phase, where half of the power interferes at the second beam splitter. Output intensities represent the case $\phi=0^\circ$. (b) Count ratio $r$ vs.{} relative \ac{mzi} phase $\phi$ in our setup. The total signal (purple) with visibility $v=1/\sqrt{8}$ is the mean value of pulse-pairs with $0^\circ$ relative phase (blue) and $90^\circ$ (red), each with a visibility of 50\,\% due to non-interfering pulses. We use two different locking points (purple dots) for the two \acp{mzi} with $r_0=5/8$ and slopes of $r'_0=\pm1/8$ at $\phi_0=0^\circ$ and $\phi_0=90^\circ$, respectively.}
\label{fig:phase-vs-r}
\end{figure}
\begin{figure}[b]
\centering
\includegraphics[width=12cm]{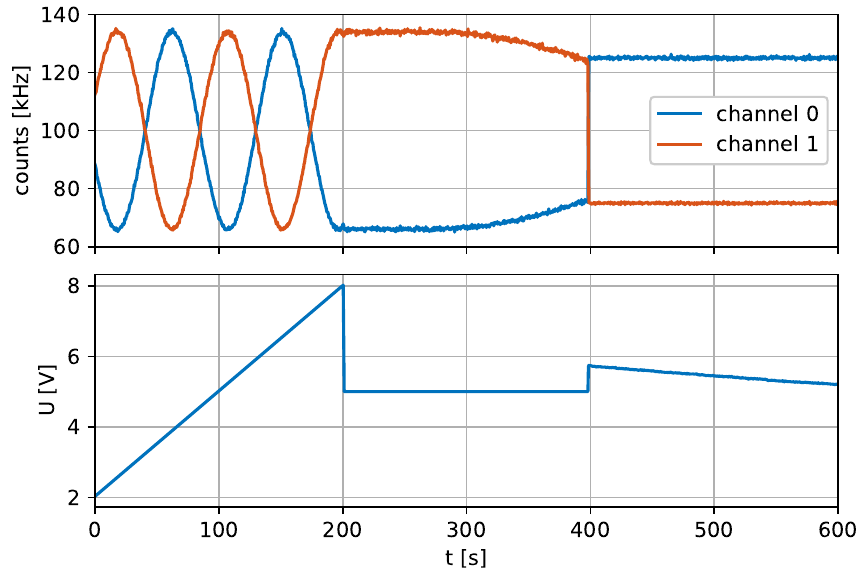}
\caption{Measured count rates after one \ac{mzi}, and stretcher voltage $U$. 0--200\,s: Linear phase sweep of $2\pi/100\,\mathrm{s}$; 200--400\,s: Free drift at $U=5\,\mathrm{V}$; 400--600\,s: Count ratio locked to $5:3$.}
\label{fig:sweep}
\end{figure}

Our fully fiber-based setup (\fref{fig:interferometer+setup}b), sketched in \fref{fig:interferometer+setup}a, consists of two identical \acp{mzi} behind a 50/50 non-polarizing beam splitter. Each \ac{mzi} decodes optical ($\lambda=1550\,\mathrm{nm}$) pulses in one independent basis, so that we can simultaneously realize two orthogonal measurement bases for \ac{qkd}. The optical input signal consists of rectangular pulse pairs with temporal separation matched by the interferometer arm length difference, with much less than one photon per pulse. The pulse pairs alternate between two different relative phases (\fref{fig:phase-vs-r}a), which enable the locking of our two \acp{mzi} to the two different phases with only one input signal. Each interferometer has its individual relative phase between its two arms, that defines its measurement basis. For \ac{qkd} operation, these pulses act as phase reference, where much weaker quantum signals are interleaved. The received pattern with partially interfering pulses results in a reduced visibility $v=1/\sqrt8$ of the mean signal (\fref{fig:phase-vs-r}b). In each \ac{mzi}, one arm contains a variable optical delay line with a range of 600\,ps for coarse adjustment of the interferometer delay $T$. The other arm contains a stretcher (FPS-002-L-15-PP) with a 10\,kHz bandwidth and a voltage-controlled phase delay range of $3.4$ wavelengths at a $0-10\,V$ input. \Fref{fig:sweep} demonstrates the output click ratio for various stretcher voltages.

To ensure polarization-mode-matching at the end of each \ac{mzi}, we use \ac{pm} fibers like in \cite{yanikgonul_phase_2020, svarc_sub_2023}, where alternative solutions are active stabilization \cite{yuan_continuous_2005} or the use of Faraday mirrors \cite{cho_stabilization_2009}. In the \ac{pm} setup, each component has a finite \ac{per} on the order of $-20\,\mathrm{dB}$ that may allow power to swap from the desired polarization mode to the orthogonal one, and back. This can decrease the interferometer visibility in a time-dependent fashion and due to amplitude interference, the worst-case effect increases quadratically with the number of subsequent imperfect components. We mitigated this effectively by the addition of two in-line clean-up polarizers in each arm, which remove wrong polarization components before they can interfere.

The arm-length in each interferometer is $7.08\,\mathrm{m}$, mainly determined by the un-shortened fiber leads of each component. All four output channels lead to \acp{SNSPD}, that are electrically connected to a time-to-digital converter. The individual timestamps of each detected photon are then processed by an FPGA (Kintex-7 160T), which performs the locking algorithm at a clock rate of $500\,\mathrm{kHz}$ and feeds back an analog signal to the fiber stretcher of each \ac{mzi} with 16-bit digital-to-analog conversion.

\begin{figure}
\centering
\includegraphics[width=12cm]{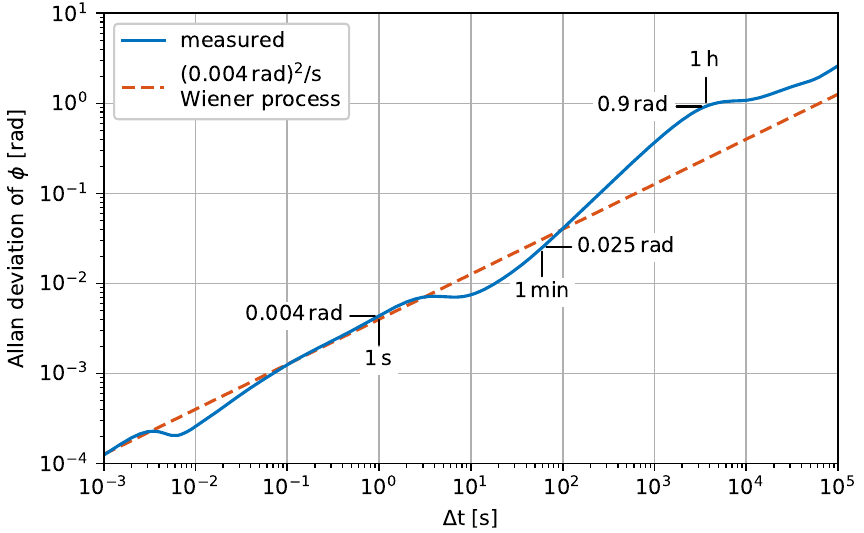}
\caption{Allan deviation of the passive phase drift of one \ac{mzi} arm with respect to the other across time intervals from $1\,\mathrm{ms}$ to $10^5\,\mathrm{s}$. On timescales from $1\,\mathrm{ms}$ to $3\,\mathrm{s}$ the drift follows approximately a Wiener process with slope 1/2 (dashed). Due to recording limitations, the progression was measured in two sections: With sampling of $f_s=2\,\mathrm{kHz}$ up to $\Delta t=1\,\mathrm{s}$, and above with $f_s=2\,\mathrm{Hz}$.}
\label{fig:allan}
\end{figure}

\begin{figure}
\centering
\includegraphics[width=12cm]{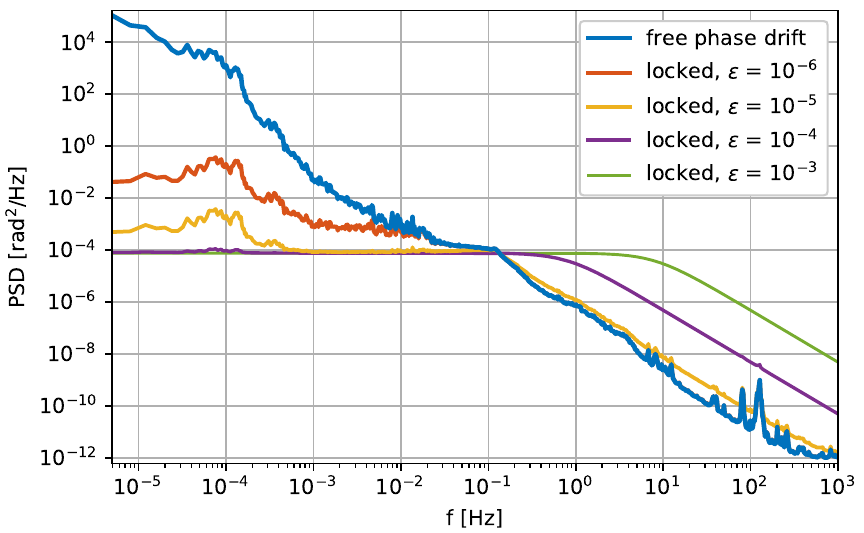}
\caption{Power spectral density (PSD) of total phase noise $S$, both free-drifting ($S_\mathrm{drift}$, blue line, measured) and locked with various locking parameters $\epsilon$ at constant count rate $f_c=200\,\mathrm{kHz}$, $r_0=5/8$ and $r'_0=1/8$, calculated via \eref{eq:spectrum-total}. At lower $\epsilon$, low-frequency drift-noise dominates, and at higher $\epsilon$, high-frequency locking-noise dominates.}
\label{fig:spectrum}
\end{figure}

The setup is passively stabilized through close contact of the fibers to a heavy metallic breadboard, mounted on four spring dampers inside of a rack drawer, lined with porous open-cell foam. The temperature is stabilized by the laboratory air conditioning. Nevertheless, the phase difference between each two \ac{mzi} arms changes naturally over time, due to mechanical stress, temperature changes and acoustic vibrations. We measured the phase drift characteristic by transmitting macroscopic (mW) light through one \ac{mzi} and observing the intensity evolution on phototiodes after the \ac{mzi} (replacing the SNSPDs in \fref{fig:interferometer+setup}a). The average amount of change over various timescales is shown in \fref{fig:allan} (time domain) and in \fref{fig:spectrum} (frequency domain). The drift characteristic follows roughly the `red noise' of a Wiener process, which is the continuous version of a random walk (dashed line in \fref{fig:allan} and slope $-2$ in \fref{fig:spectrum}). The noise amplitude at $1\,\mathrm{kHz}$ is about an order of magnitude above the expected fundamental thermal fluctuations in the fiber \cite{dong_observation_2016}. To compensate the drifts and keep the \ac{mzi} phase-difference at a constant value, we apply active feed-back through the fiber stretchers \cite{jackson_elimination_1980, fritsch_simple_1981}.

\section{Phase locking}

\subsection{Locking algorithm}
Our locking algorithm detects single-photon-detector clicks on two channels, which are the two outputs of the balanced \ac{mzi}. Let
\begin{equation}
P(0)=r\quad\mathrm{and}\quad P(1)=1-r
\end{equation}
be the relative fractions of photons received in the first and second of two channels, respectively. The ratio $r$ depends on the inter\-fero\-meter phase $\phi$ (\fref{fig:phase-vs-r}), and $r_0:=r(\phi_0)$ is the ratio for the desired phase $\phi_0$. Let $r'_0:=\frac{\rmd\,r}{\rmd\,\phi}|_{r=r_0}$ be the slope that links phase and click ratio at the locking point, which takes the magnitude $|r'_0|=\sqrt{v^2-(2r_0-1)^2}/2$ at visibility $v$.

Our regulator works in the simple manner that it changes the phase of one \ac{mzi} arm by a constant step size at each registered photon (with a negligible time delay of the FPGA clock time). The step sizes $\epsilon_0$ and $\epsilon_1$ for detections in each channel differ depending on $r_0$ and are adjusted by a step-size parameter $\epsilon$:
\begin{itemize}
    \item Photon in channel 0: $\Delta\phi = \epsilon_0 = \epsilon\cdot2(1-r_0)$
    \item Photon in channel 1: $\Delta\phi = \epsilon_1 = -\epsilon\cdot 2r_0$
\end{itemize}
Such feedback creates an average phase change at each detected photon of
\begin{equation}\label{eq:meanphasechange}\langle \Delta\phi\rangle = P(0)\cdot\epsilon_0 + P(1)\cdot\epsilon_1 = 2\,\epsilon\cdot(r-r_0)
\ .
\end{equation}
Thus, in sufficient proximity to the locking point, the average phase adjustment is
\begin{equation}\label{eq:regulator_step}
    \langle \Delta\phi\rangle = 2\,\epsilon\cdot(r-r_0) = 2\,\epsilon \cdot r'_0\cdot(\phi-\phi_0)\ ,
\end{equation}
proportional to the error of $\phi$. Therefore, we effectively integrate up the phase proportionally to its error, which constitutes an integral~(I)-regulator.

As the step-size is small, we can express the differential progression in time at total photon count rate $f_c$ as
\begin{equation}\label{eq:regulator_dif}
   \frac{\langle \rmd\phi\rangle}{\rmd t} = \langle \Delta\phi\rangle \cdot f_c = 2\,\epsilon \cdot r'_0\cdot(\phi-\phi_0)\cdot f_c\ ,
\end{equation}
which causes exponential damping of phase errors in time
\begin{equation}\label{eq:regulator_exp}
   \phi(t) = \phi_0 + \phi(t{=}0)\cdot \rme^{-\theta\cdot t}\ ,
\end{equation}
with regulator stiffness (exponential decay rate)
\begin{equation}
    \theta = -2\,\epsilon\,r'_0 f_c\ ,
\end{equation}
time constant
\begin{equation}
    \tau = \frac{1}{\theta} = \frac{1}{-2\,\epsilon\,r'_0 \, f_c}\ ,
\end{equation}
and locking bandwidth
\begin{equation}\label{eq:flock}
f_{\mathrm{lock}} = \frac{\theta}{2\pi} = \frac{-\epsilon\,r'_0 \, f_c}{\pi}\ .
\end{equation}

\subsection{Discrete locking noise}
In addition to the linear feedback, there is stochastic noise from the random nature of the photon statistics (Poissonian in time and binomial per detection). Each detection is a Bernoulli trial, and the phase variance increases by the variance $V$ of a binomial distribution with probability $r$, which is
\begin{eqnarray}\label{eq:variance-lock}
    V = \sum_{i\in\{0,1\}} P(i)\cdot(\epsilon_i - \langle \Delta\phi\rangle)^2 = 4\epsilon^2 r(1-r)\ .
\end{eqnarray}

The successive phase adjustments create a phase random-walk with mean step $\langle \Delta\phi\rangle$ and an added variance per step $V$. For sufficiently small phase errors ($|\phi-\phi_0|\ll\pi/2$) which we find in the experiment, and thus $r$ close to $r_0$ and a near-constant slope $r'_0$, the variance can be approximated by the constant value $V=4\epsilon^2 r_0(1-r_0)$. At small step sizes, the phase evolution follows the stochastic differential equation
\begin{equation}
    \rmd\phi = -\theta\cdot(\phi-\phi_0)\,\rmd t + \sigma\,\rmd W_t
\end{equation}
where $W_t$ is a Wiener process, $\sigma=\sqrt{Vf_c}$, and the diffusion constant is $D=Vf_c/2$. Such a random-walk with linear feedback is called an \textbf{\ac{ou} process} with stiffness $\theta$ and diffusion $\sigma$ \cite{wang_theory_1945, gillespie_exact_1996}. This has not been previously identified in the context of phase-locking, and provides the basis for a deep understanding of the locking dynamic. It follows that the probability distribution of phases around the desired phase $\phi_0$ is Gaussian with a standard deviation of
\begin{equation}\label{eq:sigma_ou}
        \sigma_{\phi,\mathrm{lock}} = \sqrt{\frac{D}{\theta}} = \sqrt{\frac{\epsilon \,r_0(1-r_0)}{-r'_0}}
        \ .
\end{equation}
Here, in order for $\phi_0$ to be a stable locking point, $\epsilon$ and $r'_0$ need to have opposite signs. It is evident from equation~(\ref{eq:sigma_ou}) that in absence of external noise, the locking error scales with the square root of the chosen step size $|\epsilon|$.
\begin{figure}
\centering
\includegraphics[width=12cm]{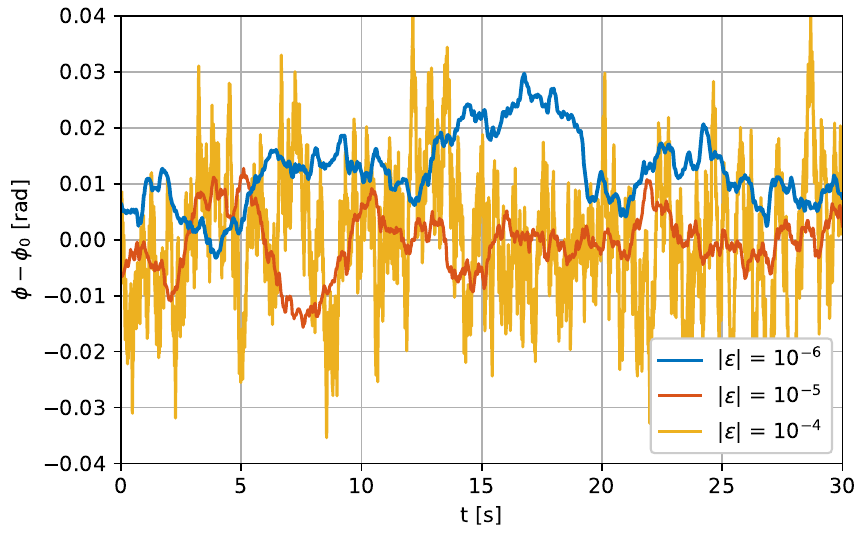}
\caption{Sample traces of the phase error during locking for step sizes of $|\epsilon|=10^{-6}$, $|\epsilon|=10^{-5}$ and $|\epsilon|=10^{-4}$, with locking time constants of $\tau=20\,\mathrm{s}$, $\tau=2\,\mathrm{s}$ and $\tau=0.2\,\mathrm{s}$, respectively.}
\label{fig:phasetrace}
\end{figure}

The (two-sided) power-spectral-density of the phase progression is that of low-pass filtered white noise with a cutoff frequency of $f_\mathrm{lock}$ \cite{wang_theory_1945, bibbona_ornsteinuhlenbeck_2008}
\begin{equation}\label{eq:ou-spectrum}
S_\mathrm{lock}(f) = \frac{r_0(1-r_0)}{r'^2_0\,f_c(1 + (f/f_\mathrm{lock})^2)}\ ,
\end{equation}
where $\int_{-\infty}^{\infty}S_\mathrm{lock}(f)\,\rmd f = \sigma_{\phi,\mathrm{lock}}^2$.

\subsection{Total phase error}
\begin{figure}
\centering
\includegraphics[width=12cm]{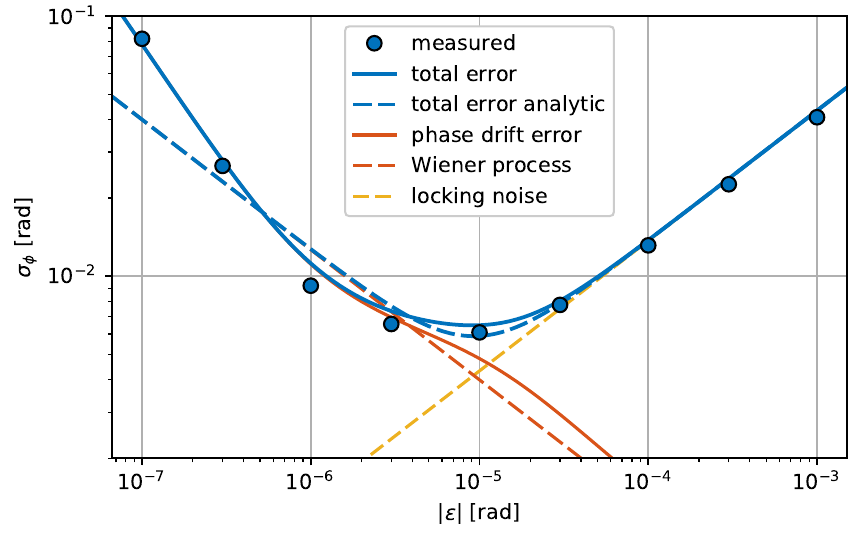}
\caption{Phase error vs{.} locking step-size parameter $|\epsilon|$ without external drift at $f_c=200\,\mathrm{kHz}$, $r_0=5/8$ and $r'_0=1/8$. Solid lines are computed from \eref{eq:sigma-integrated} with the measured drift spectrum of \fref{fig:spectrum}, and dashed lines from the analytic model of (\ref{eq:sigma_wiener}).}
\label{fig:err_vs_epsilon}
\end{figure}
The total phase error is a combination of the locking error and the residual phase drift. Due to independence (locking noise is random), both variances add up, and so do their power spectra. Like every I-regulator, the lock is basically a first-order high-pass filter on the free phase drift (of spectrum $S_\mathrm{drift}$, \fref{fig:spectrum}) with cutoff-frequency $f_{\mathrm{lock}}$ \eref{eq:flock}. In addition, the locking noise (\ref{eq:sigma_ou}, \ref{eq:ou-spectrum}) is added. This is most simply expressed in the spectral domain, where the total noise spectrum becomes
\begin{equation}\label{eq:spectrum-total}
S(f) = S_\mathrm{lock}(f) + \frac{S_\mathrm{drift}(f)}{1 + (f_{\mathrm{lock}} / f)^2}\ ,
\end{equation}
and the total phase error
\begin{equation}\label{eq:sigma-integrated}
    \sigma_{\phi} = \sqrt{\int_{-\infty}^{\infty}S(f)\,\rmd f} = \sqrt{\sigma_{\phi,\mathrm{lock}}^2 + \int_{-\infty}^{\infty}\frac{S_\mathrm{drift}(f)}{1 + (f_{\mathrm{lock}} / f)^2}\,\rmd f}\ .
\end{equation}
At larger step-sizes $|\epsilon|$, the free phase drift gets suppressed more and with a higher cutoff-frequency, but the locking-noise increases in bandwidth and magnitude (\fref{fig:spectrum}). Therefore we can find an optimum magnitude of $\epsilon$, for which the total noise is minimal.

\Fref{fig:err_vs_epsilon} shows this dependence for a fixed count-rate. The experimental phase noise for this (\fref{fig:phasetrace}) was measured with macroscopic optical power on photodiodes and artificially sampled Poissonian photon counts for locking. Measured error values $\sigma_\phi$ follow the predictions with a slight variation, due to the fiber phase drift behaviour changing gradually over the measurement time of the spectrum of several weeks (for $\mathrm{\mu Hz}$ frequency components), as the setup relaxed.

\subsubsection{Linear phase drift approximation}
Let us now analyze the behaviour for linear phase drifts. Such drifts occur for instance when the fiber temperature changes continuously. In \fref{fig:allan}, they appear on timescales between 50\,s and 2000\,s, where the phase changes proportional to $\Delta t$. Linear drifts are also induced from changing Doppler shifts in satellite \ac{qkd}, where the maximum frequency chirp from low Earth orbits at altitude $h$ and speed $v_o$ is $\gamma = \rmd f/\rmd t = v_o^2/(h\,\lambda) \approx c/\lambda\cdot4\cdot 10^{-7}\,\mathrm{s^{-1}}$. In a \ac{mzi} of path difference $T$, this induces a phase drift of $d=\rmd \phi/\rmd t=2\pi\gamma T$ on the order of $0.08\,\mathrm{rad/s}$.

At a mean photon count rate $f_c$ per \ac{mzi}, a phase step size $\epsilon$ and a locking ratio $r_0$ at phase $\phi_0$, the average phase drift during each count is $\Delta\phi_\mathrm{drift} = d/f_c$. The equilibrium is reached when the drift becomes opposite equal to the mean locking correction $\langle\Delta\phi\rangle$, thus
    \begin{equation}
        \Delta\phi_\mathrm{drift} = -\langle\Delta\phi\rangle\ ,
\qquad
        \frac{d}{f_c} = -2\,\epsilon \, r'_0\cdot(\phi-\phi_0)\ ,
    \end{equation}
therefore
    \begin{equation}\label{eq:drift_error}
        \phi_\mathrm{drift} = \phi-\phi_0 = -\frac{d}{2f_c\,\epsilon\,r'_0}\ .
    \end{equation}
This drift error is proportional to $1/\epsilon$. Together with the locking error (\ref{eq:sigma_ou}), it leads to a total phase error of
    \begin{equation}\label{eq:drift_error_total}
        \sigma_{\phi,\mathrm{drift}} = \sqrt{\phi_\mathrm{drift}^2 + \sigma_{\phi,\mathrm{lock}}^2} = \sqrt{\frac{d^2}{(2f_c\,\epsilon\,r'_0)^2}+\frac{\epsilon \,r_0(1-r_0)}{-r'_0}}\ ,
    \end{equation}
which takes a minimum value
\begin{equation}\label{eq:drift-optimum-sigma}
    \min(\sigma_{\phi, \mathrm{drift}}) = \sqrt{3} \sqrt[3]{\frac{|d|\, r_0(1-r_0)}{4f_c r'^2_0}}
\end{equation}
at an optimum stepsize
\begin{equation}\label{eq:drift-optimum-eps}
    \epsilon_{\mathrm{opt}, \mathrm{drift}} = \sqrt[3]{\frac{d^2}{2f_c^2r_0(1-r_0)|r'_0|}}\mathop{\textrm{sign}}(-r'_0)
    \ .
\end{equation}

\subsubsection{Wiener phase drift approximation}
On timescales below 3\,s, at which the locking typically operates, the free phase drift (\fref{fig:allan}) is roughly proportional to $\Delta t^{1/2}$, a Wiener process of random phase drifts. For this simplified case, we can again estimate the locking behaviour analytically. The diffusion constant in our case is $D_\mathrm{fiber}=(4\,\mathrm{mrad})^2/\mathrm{s}$. This type of phase-noise can be easily included in the variance of the locking \ac{ou} process from equation~(\ref{eq:variance-lock}) as
\begin{equation}
    V = 4\epsilon^2 r(1-r) + D_\mathrm{fiber}/f_c
\end{equation}
to yield a total phase error (analogous to equation~(\ref{eq:sigma_ou})) of
\begin{equation}\label{eq:sigma_wiener}
    \sigma_{\phi, \mathrm{Wiener}} = \sqrt{\frac{4\epsilon^2 \,r_0(1-r_0) + D_\mathrm{fiber}/f_c}{-4\epsilon r'_0}}\ ,
\end{equation}
which takes a minimum value
\begin{equation}\label{eq:Wiener-optimum-sigma}
    \min(\sigma_{\phi, \mathrm{Wiener}}) = \sqrt[4]{\frac{D_\mathrm{fiber}\, r_0(1-r_0)}{f_c r'^2_0}}
\end{equation}
at an optimum stepsize
\begin{equation}\label{eq:Wiener-optimum-eps}
    \epsilon_{\mathrm{opt}, \mathrm{Wiener}} = \sqrt{\frac{D_\mathrm{fiber}}{4f_cr_0(1-r_0)}}\mathop{\textrm{sign}}(-r'_0)
    \ .
\end{equation}

\subsection{Count rate dependence}
\begin{figure}
\centering
\includegraphics[width=12cm]{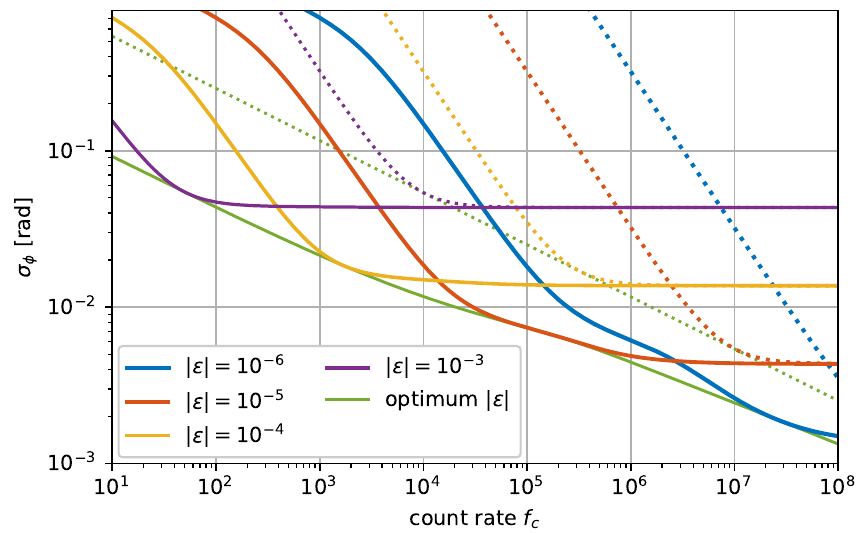}
\caption{Phase error versus count rate $f_c$ for various fixed values of the locking parameter $|\epsilon|$. Curves are computed with \eref{eq:sigma-integrated} using the measured free drift spectrum, $r_0=5/8$ and $r'_0=1/8$. Solid lines are without external phase drift, dotted lines with linear external drift of $d=0.08\,\mathrm{rad/s}$.}
\label{fig:err_vs_countrate}
\end{figure}

\Fref{fig:err_vs_countrate} shows the achievable root-mean-squared phase error $\sigma_\phi$ versus received photon count rates $f_c$. The locking generally improves with larger $f_c$, as the available information increases. Larger locking step sizes $|\epsilon|$ lead to better noise suppression at smaller count rates, because the lock will act stronger against phase deviations. However, a larger $|\epsilon|$ also leads to more locking noise, that dominates at higher count rates. Therefore, as in (\ref{eq:drift-optimum-eps}) and (\ref{eq:Wiener-optimum-eps}), the optimum $\epsilon$ depends on $f_c$.

In absence of external phase drifts, $|\epsilon|=10^{-5}\,\mathrm{rad}$ is a near-optimum choice in our setup for a wide range of count rates from around $3\cdot 10^4\,\mathrm{Hz}$ to $10^6\,\mathrm{Hz}$. For the default count rate $f_c=200\,\mathrm{kHz}$, this yields $\sigma_{\phi, \mathrm{min}}=6.5\,\mathrm{mrad}$ with a locking bandwidth of $f_\mathrm{lock}=0.08\,\mathrm{Hz}$.

At presence of external phase drifts (dotted lines in \fref{fig:err_vs_countrate}), the required count rate to suppress drift errors generally increases. This can be mitigated by a larger $|\epsilon|$ at the cost of increased minimum achievable phase accuracy. For instance, a desired phase accuracy $\Delta\phi=\pi/100$ allows for a maximum $|\epsilon|=5\cdot10^{-4}\,\mathrm{rad}$ (\ref{eq:sigma_ou}). The desired accuracy can then be maintained down to $f_c=1\,\mathrm{kHz}$, where the locking bandwidth reduces to $f_\mathrm{lock}=0.02\,\mathrm{Hz}$. For a minimal phase error over a wider range of count rates, it can make sense to choose $\epsilon$ adaptively to the count rate, following the green lines in \fref{fig:err_vs_countrate}).

\subsection{Darkcounts}
The effect of darkcounts (or random-phase quantum signals) is most simply included by a reduced visibility $v$ of $r(\phi)$, because darkcounts are constant with regard to the \ac{mzi} phase. Therefore, darkcounts may shift the locking point $r_0$ (towards 1/2 if the darkcounts are equal in both channels) and they flatten the slope $r'_0$ by a factor of $f_{\mathrm{dark}}/f_{\mathrm{total}}$. In practice, with dark count rates of few Hz and signal count rates of hundreds of kHz, the effect is often negligible.

\subsection{Optimality of the direct-counting I-controller}
Our locking scheme of applying immediate constant phase changes at each registered photon is not just simple, but also optimal with regard to some often-used modifications:

First, applying immediate feedback is better than additional averaging over several counts $n$ (as for example applied in \cite{pulford_single_2005, yanikgonul_phase_2020}).
When averaging over subsequent counts, the mean stepsize from equation~(\ref{eq:meanphasechange}) becomes $\langle \Delta\phi_n\rangle = n\cdot2\,\epsilon\, r'_0\cdot(\phi-\phi_0)$, and $\theta_n = -2\,n\,\epsilon r'_0 (f_c/n) = -2\,\epsilon r'_0 f_c = \theta$ for any chosen $\epsilon$. The added variance at each phase adjustment from equation~(\ref{eq:variance-lock}) becomes $V_n = n\cdot4\epsilon^2 r(1-r)$, because independent variances add up, and thus $D_n=V_n(f_c/n)/2=D$. Together this yields $ \sigma_{\phi,\mathrm{lock},n} = \sqrt{D_n/\theta_n} = \sigma_{\phi,\mathrm{lock}}$, the same locking noise as without averaging. The only difference is an additional mean time delay of $n/(2f_c)$ in the feedback, which will slow down the locking response and degrade the suppression of external phase-noise. Therefore, it is best to adjust the phase immediately on each detection of a single photon.

Second, instead of pure integral~(I)-regulation, a PID-controller with nonzero proportional (P) or differential (D) parts might be employed (for example, PI-control in \cite{pulford_single_2005}). The I-part is required in order to accumulate long-term phase drifts. The advantage of a PI-controller over a pure I-controller is that it is normally faster, because the P-part can react immediately, while the I-part needs to integrate for a time that is longer than the actuator and sensor loop delay (few $100\,\mathrm{\mu s}$ here) to avoid oscillation. In the low-count-rate regime, however, immediate P-response is impossible, because low-noise statistics on the discrete count ratios is only acquired on timescales much longer than the loop delay. Then, however, the I-part has already utilized the corresponding clicks and the P-response comes too late to add anything useful. Things are even worse for a D-part, as the differentiation makes it even more prone to noise, and the required integration time would be even longer.

\section{Conclusion}
We have laid out and implemented a novel phase-locking scheme for \acp{mzi}, that utilizes discrete detections of single photons. As demonstrated, immediate feedback on each detected photon is optimal in the low count-rate regime. Despite the limitation of a relatively low locking frequency in the Hz-range, inherently restricted by the available information, we were able to achieve a very low phase error of $6\,\mathrm{mrad}$ ($0.35^\circ$) in our $7\,\mathrm{m}$-long interferometer.

Our method is very hardware efficient. In contrast to systems with separate locking light, it requires no additional lasers, modulators, filters or detectors in addition to the signal laser and detectors. The interferometer can be locked to any phase value where the slope of click ratios $r'_0$ is nonzero. In case of an initially vanishing slope, the desired phase value can be made accessible by the injection of pulse pairs with carefully chosen relative phases, as we have demonstrated.

The simplicity of our scheme, to move a fixed phase step on arrival of every photon, allows it to be implemented straightforwardly on basic hardware, that does not necessarily include an FPGA. It can function with any type of single-photon detector, such that instead of the SNSPD, more affordable devices like photomultipliers or avalanche diodes can be equally employed. The locking scheme may find applications in various optical interferometers at low intensities, such as quantum key distribution setups \cite{yuan_continuous_2005, dynes_stability_2012}, quantum repeaters \cite{minar_phase-noise_2008}, precision measurements \cite{kirkendall_overview_2004} and receivers for deep-space probes \cite{zwolinski_range_2018, mohageg_deep_2022}.

\ack{Part of this research was carried out within the scope of the QuNET project, funded by the German Federal Ministry of Education and Research (BMBF) in the context of the federal government’s research framework in IT security `Digital. Secure. Sovereign.'. The authors are grateful for financial support from the Bavarian State Ministry of Economic Affairs and Media, Energy and Technology through the project `Satellitengestützte Quantenkryptografie' BayernQSat (LABAY98A).}

\section*{References}
\providecommand{\newblock}{}

\end{document}